\documentclass{moriond}

\def\be{\begin{equation}}
\def\ee{\end{equation}}
\def\bea{\begin{eqnarray}}
\def\eea{\end{eqnarray}}

\newcommand{\Photo}{}

\begin{document}
\vspace*{4cm}
\title{Multi-messenger Constraints on a Primordial Black Hole Origin of the KM3-230213A Event}

\author{Yuber F. Perez-Gonzalez}

\address{Departamento de F\'{i}sica Te\'{o}rica and Instituto de F\'{i}sica Te\'{o}rica (IFT) UAM/CSIC, Universidad Aut\'{o}noma de Madrid, Cantoblanco, 28049 Madrid, Spain}

\maketitle\abstracts{
    Black holes are expected to end their lifetime in a burst of Hawking radiation, emitting all Standard Model particles at ultra-high energies. The evaporation of a nearby primordial black hole (PBH) has been proposed as an explanation for the high-energy neutrino-like event reported by KM3NeT. Such a scenario requires the source to be extremely close to Earth, implying detectable gamma-ray and cosmic-ray emission. Accounting for the time-dependent field of view of gamma-ray observatories, we find that current experiments should have observed a pre-burst signal, while neutrino telescopes would also detect lower-energy events before the burst. The absence of such multimessenger signals strongly disfavors a PBH origin of the KM3-230213A event in the minimal 4D Schwarzschild scenario.
}

\section{Introduction}

The detection of an ultra-high-energy neutrino event by the KM3NeT Collaboration~\cite{KM3NeT:2025npi} has generated significant interest in addressing its origin. With an energy of $\mathcal{O}(100)\,\mathrm{PeV}$ and no clear counterpart in experiments such as IceCube, it challenges standard interpretations based on diffuse or steady sources, motivating transient scenarios.

One possibility is that the event originates from the final burst of a primordial black hole (PBH) undergoing Hawking evaporation~\cite{Klipfel:2025jql}. In this case, a nearby PBH could produce high-energy neutrinos compatible with the observation. However, this scenario also predicts accompanying multimessenger signals that have not been observed~\cite{Airoldi:2025opo,Airoldi:2025bgr}. In this work, we assess the viability of a PBH origin for the KM3NeT event.

\section{Primordial Black Holes}

Astrophysical black holes form from the collapse of massive stars and are bounded by the Chandrasekhar limit, of order one solar mass. The formation of lighter black holes therefore requires alternative conditions, such as those operating in the Early Universe. Primordial black holes (PBHs) can be produced through several mechanisms, the most studied being the collapse of large density fluctuations upon horizon re-entry. They may also constitute a fraction of the dark matter (DM); see e.g.~Ref.~\cite{Carr:2026hot} for a review. 

\subsection{Black Hole evaporation}

Black holes emit particles due to the non-trivial evolution of spacetime during gravitational collapse, producing a particle flux that follows a thermal spectrum. The spectrum of particles $j$, with internal degrees of freedom $g_j$ and spin $s_j$, emitted from non-rotating chargeless Schwarzschild black holes is 
\begin{eqnarray}
    \frac{d^2 N_j}{d \omega d t} = \frac{g_j}{2\pi} \frac{\Gamma_{s_j}(\omega)}{\exp(\omega/T) - (-1)^{2s_j}}\, ,
\end{eqnarray}
where $T$ is the black hole temperature and $\omega$ the particle energy. The factors $\Gamma_{s_j}(\omega)$, known as graybody factors, encode deviations from a purely Planckian spectrum due to the surrounding curved spacetime acting like a potential barrier. 
The BH temperature for a Schwarzschild BH with a mass $M$ is
\begin{equation}
    T =\frac{1}{8\pi GM} \sim 1~{\rm GeV}\left(\frac{10^{13}~{\rm g}}{M}\right),
\end{equation}
Thus, more massive BHs have lower temperatures than lighter ones. This distinctive dependence on the BH mass will be crucial for assessing whether the KM3NeT event can be attributed to a PBH origin. 

Among the emitted particles, neutrinos and photons are particularly relevant, as they can be detected by current experiments. They arise both directly from Hawking radiation (primary component) and from the decay of unstable particles (secondary component). We model the latter using \texttt{HDMSpectra}~\cite{Bauer:2020jay} as implemented in \texttt{BlackHawk}~\cite{Arbey:2021mbl}. As an illustration, Fig.~\ref{fig:integrated_spectra} (left) shows the emission spectra for neutrinos and photons from a PBH with $M = 2.5 \times 10^5~\mathrm{g}$. The dashed lines indicate the primary contribution, while the shaded region corresponds to the KM3NeT energy range. This highlights that very light black holes are required to reach such high energies.

\subsection{Time evolution}

Energy conservation implies that black holes lose mass according to\footnote{This evolution equation involves some caveats. In particular, the semiclassical approximation breaks down near the Planck scale, and the information problem may modify the evolution. See, e.g., Ref.~\cite{Perez-Gonzalez:2025try} for details.}
\begin{eqnarray}\label{eq:dMdt}
    \frac{d M}{d t} = - \sum_j \int dE\, E \frac{d^2 N_j}{dE dt} \equiv -\frac{\varepsilon(M)}{G^2 M^2}.
\end{eqnarray}
The evaporation is initially slow but accelerates, leading to a runaway evolution and a final burst in which the emitted particles reach very high energies.
\begin{figure}
    \centering
    \includegraphics[width=0.4\linewidth]{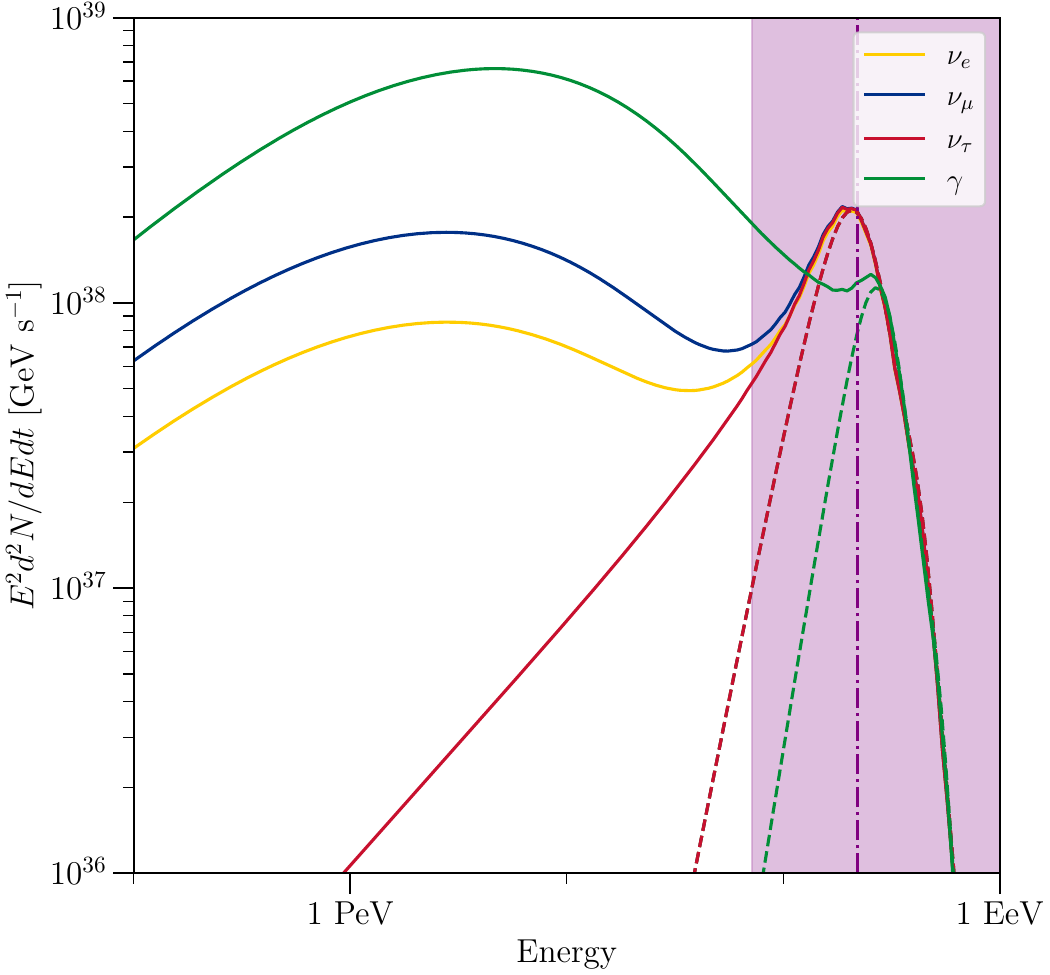}
    \includegraphics[width=0.4\linewidth]{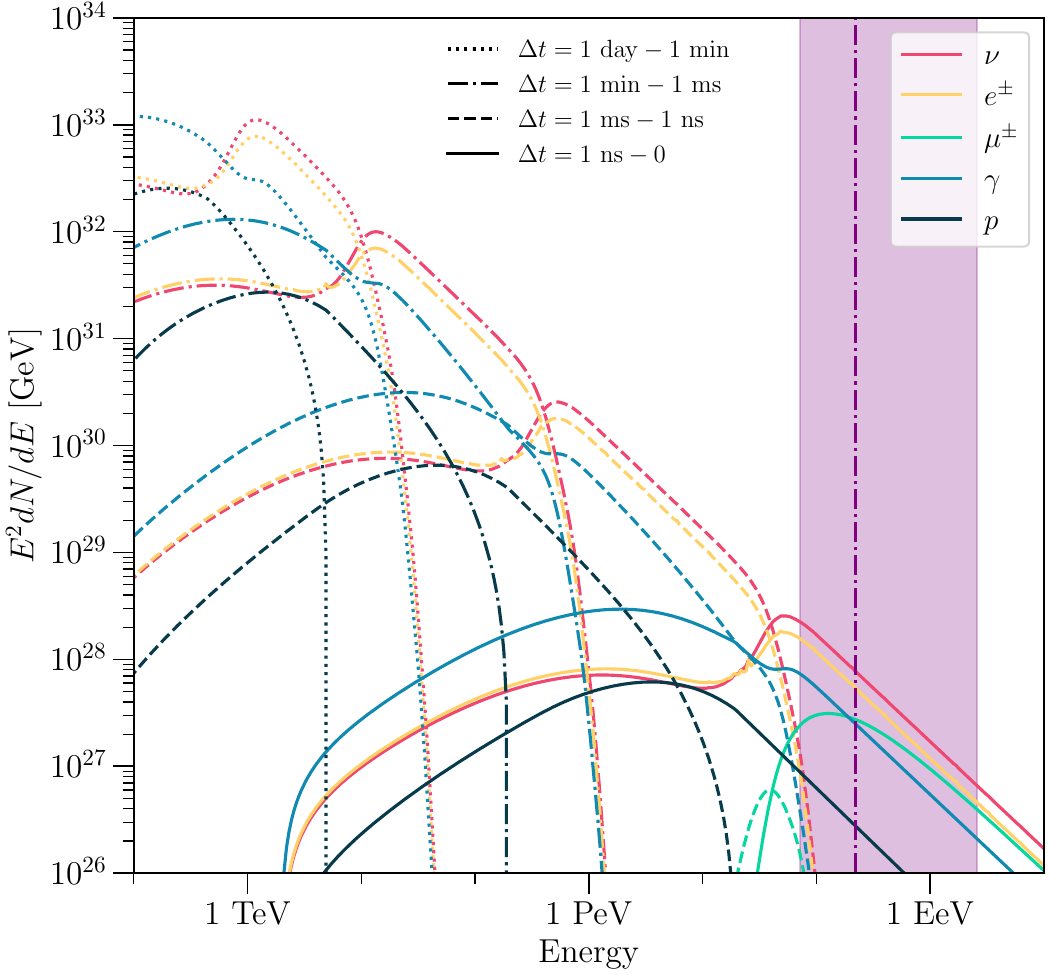}
    \caption{Left: Hawking spectra for neutrinos and photons from a PBH with $M=2.5\times10^5~{\rm g}$ (dashed: primary). Right: time-integrated spectra for different intervals before evaporation. The purple band indicates the KM3-230213A energy range (90\% C.L.).}
    \label{fig:integrated_spectra}
\end{figure}

To study the spectra close to evaporation, we compute
\begin{eqnarray}
   \frac{dN_j}{dE} = \int_{\tau - \Delta t}^{\tau} dt\, \frac{d^2 N_{j}}{dE dt}(M(t)),
\end{eqnarray}
where $\tau$ is the PBH lifetime and $\Delta t$ a time interval before evaporation. Fig.~\ref{fig:integrated_spectra} (right) shows the time-integrated spectra for different particle species and time intervals before evaporation. For muons, we assume a distance of $10^{-5}\,\mathrm{pc}$ between the PBH and the observation point, and we consider only Standard Model degrees of freedom.

\section{KM3-230213A as a Primordial Black Hole burst}

Having established that PBHs can emit particles at energies comparable to the KM3NeT event, we estimate the expected signal in neutrino and gamma-ray detectors. In general,
\begin{eqnarray}\label{eq:Nevts}
    N_{\rm evt}^j = \int dE \, \frac{d \Phi_{j}}{d E}\,  A_{\rm eff}(E,\delta_{\rm PBH},{\rm RA}_{\rm PBH}),
\end{eqnarray}
where $\frac{d \Phi_{j}}{d E}$ is the particle flux and $A_{\rm eff}$ the detector effective area in the event direction parametrized via celestial coordinates $(\delta_{\rm PBH},{\rm RA}_{\rm PBH})$. We consider two scenarios: a diffuse PBH population and a single nearby source. 

\subsection{Diffuse PBH population} 

Let us first consider a PBH population with a monochromatic initial mass distribution, such that the black holes evaporate at the time of the KM3NeT event. The resulting neutrino flux can be written as \begin{eqnarray}\label{eq:flux_diffuse_mono} 
    \frac{d \Phi_{\nu}}{d E} = \frac{1}{4\pi} D(\Delta \Omega)\, \frac{f_{\rm PBH}}{M_*}\, \frac{dN_j}{dE}, 
\end{eqnarray} 
where $D(\Delta \Omega) = \int_{\Delta\Omega} d\Omega \int ds\, \rho\big(r(s,l,b)\big)$ is the standard $D$-factor, $\rho(r)$ is the Galactic dark matter density (assumed to follow a Navarro-Frenk-White profile), and $\Delta\Omega$ denotes the region of the sky compatible with the KM3NeT event direction. The parameter $f_{\rm PBH}$ represents the fraction of dark matter in PBHs, and $M_* \approx 5.4 \times 10^{14}~\mathrm{g}$ is the mass of a PBH with a lifetime equal to the age of the Universe. 

To assess whether this scenario can account for the observed event, we determine the value of $f_{\rm PBH}$ required to yield one detected event. Using the KM3NeT effective areas described in Ref.~\cite{Airoldi:2025opo}, we obtain $f_{\rm PBH} \approx 2\times10^{-7} - 2\times10^{-6}$. These values are in tension with existing constraints (see Ref.~\cite{Carr:2026hot}), thereby excluding the possibility that the KM3NeT event originates from a diffuse PBH population evaporating simultaneously. 

\subsection{A nearby black hole} 

A second possibility is the presence of a single nearby PBH. In this case, the neutrino flux is given by 
\begin{eqnarray}
\label{eq:flux_nearby} 
\frac{d \Phi_{\nu}}{d E} = \frac{1}{4\pi d_{\rm PBH}^2} \frac{dN_j}{dE}, 
\end{eqnarray} 
where $d_{\rm PBH}$ denotes the distance between the PBH and Earth, and the time interval $\Delta t$ corresponds to the KM3NeT livetime. This scenario was first considered in Ref.~\cite{Klipfel:2025jql}. To assess its viability, we determine the value of $d_{\rm PBH}$ required to reproduce the observed event and study the corresponding implications for other experiments. Using the KM3NeT effective areas as described in Ref.~\cite{Airoldi:2025opo}, we find $d_{\rm PBH} \approx (1\!-\!7)\times10^{-5}\,\mathrm{pc}$. Such a PBH would therefore lie within the Solar System at the time of the final burst, implying a potentially detectable flux of high-energy particles in gamma-ray observatories. 

\begin{figure} 
\centering 
\includegraphics[width=\linewidth]{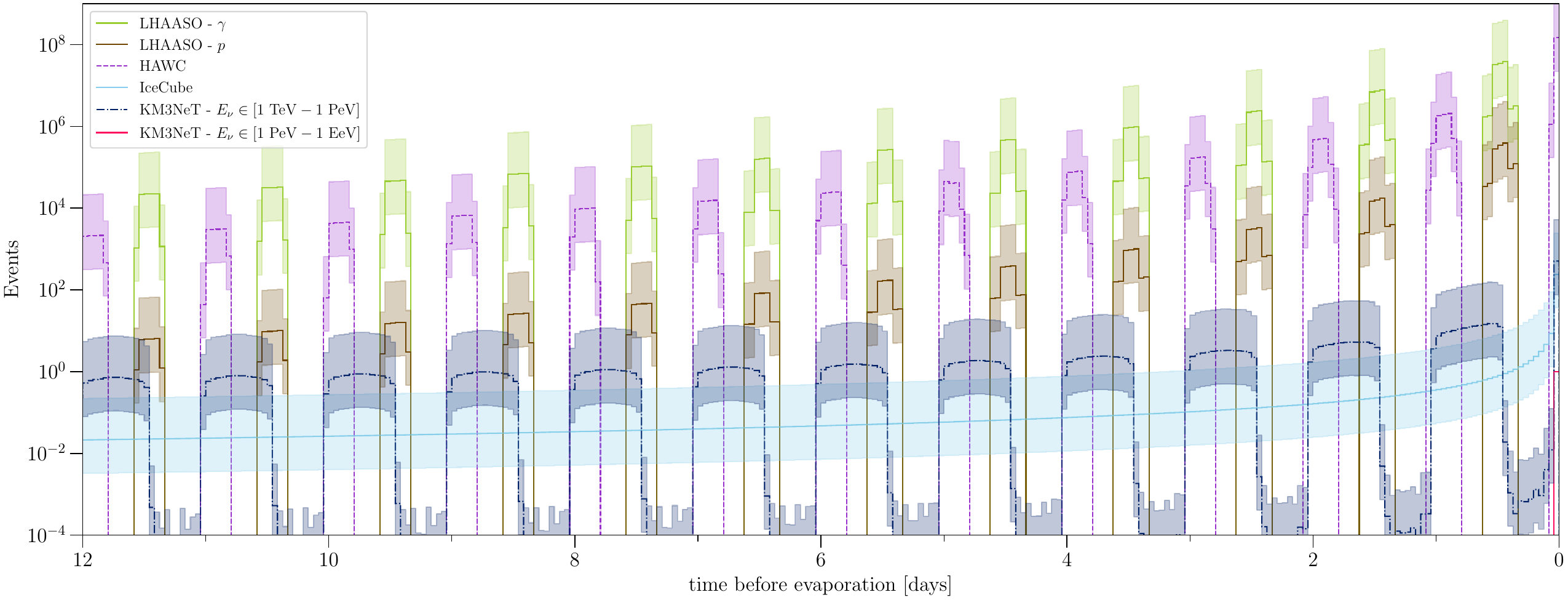} 
\caption{Time evolution of expected events in neutrino and gamma ray/cosmic rays detectors twelve days prior to the KM3NeT event, assumed it to have a PBH origin. We present events in LHAASO ($\gamma$ - green, $p$ - brown), HAWK (dashed purple), IceCube (light blue), KM3NeT low energies ($E_\nu \in [1~\mathrm{TeV},\,1~\mathrm{PeV}]$ dark blue)} 
\label{fig:events_12day} 
\end{figure}

Taking into account the time-dependent field of view of experiments such as LHAASO and HAWC, Fig.~\ref{fig:events_12day} shows the time evolution of the expected signals in these detectors, as well as in IceCube and KM3NeT in the energy range $E_\nu \in [1~\mathrm{TeV},\,1~\mathrm{PeV}]$, up to twelve days before evaporation. We find that HAWC should have detected the final burst, while LHAASO would have observed a large number of events in the hours preceding it, as well as a significant signal at earlier times. The absence of such signatures strongly disfavors a PBH origin of the KM3NeT event in the minimal scenario.

\section*{Acknowledgments}

The author would like to thank the participants of the Rencontres de Moriond session on Very High Energy Phenomena in the Universe for insightful questions that motivated the determination of the PBH dark matter fraction required to explain the KM3NeT event in the diffuse PBH population scenario.

\section*{References}

\begin{thebibliography}{99}

\bibitem{KM3NeT:2025npi}
S.~Aiello \textit{et al.} [KM3NeT],
``Observation of an ultra-high-energy cosmic neutrino with KM3NeT,''
Nature \textbf{638} (2025) no.8050, 376-382
[erratum: Nature \textbf{640} (2025), E3]

\bibitem{Klipfel:2025jql}
A.~P.~Klipfel and D.~I.~Kaiser,
``Ultrahigh-Energy Neutrinos from Primordial Black Holes,''
Phys. Rev. Lett. \textbf{135} (2025) no.12, 121003
[arXiv:2503.19227 [hep-ph]].

\bibitem{Airoldi:2025opo}
L.~F.~T.~Airoldi, G.~F.~S.~Alves, Y.~F.~Perez-Gonzalez, G.~M.~Salla and R.~Z.~Funchal,
``Could a Primordial Black Hole Explosion Explain the Extremely High-Energy KM3NeT Neutrino Event?,''
Phys. Rev. Lett. \textbf{136} (2026) no.4, 041002
[arXiv:2505.24666 [hep-ph]].

\bibitem{Airoldi:2025bgr}
L.~F.~T.~Airoldi, G.~F.~S.~Alves, Y.~F.~Perez-Gonzalez, G.~M.~Salla and R.~Z.~Funchal,
``Tackling transient sources with neutrino telescopes,''
Phys. Rev. D \textbf{113} (2026) no.2, 023052
[arXiv:2505.24652 [astro-ph.HE]].

\bibitem{Carr:2026hot}
B.~Carr, A.~J.~Iovino, G.~Perna, V.~Vaskonen and H.~Veerm{\"a}e,
``Primordial black holes: constraints, potential evidence and prospects,''
[arXiv:2601.06024 [astro-ph.CO]].

\bibitem{Bauer:2020jay}
C.~W.~Bauer, N.~L.~Rodd and B.~R.~Webber,
``Dark matter spectra from the electroweak to the Planck scale,''
JHEP \textbf{06} (2021), 121
[arXiv:2007.15001 [hep-ph]].

\bibitem{Arbey:2021mbl}
A.~Arbey and J.~Auffinger,
``Physics Beyond the Standard Model with BlackHawk v2.0,''
Eur. Phys. J. C \textbf{81} (2021), 10
[arXiv:2108.02737 [gr-qc]].

\bibitem{Perez-Gonzalez:2025try}
Y.~F.~Perez-Gonzalez,
``Page time of primordial black holes in the Standard Model and beyond,''
Phys. Rev. D \textbf{111} (2025) no.8, 083015
[arXiv:2502.04430 [astro-ph.CO]].



\end{thebibliography}


\end{document}